\documentclass[aip,apl,reprint,preprintnumbers]{revtex4-1}
\usepackage{graphicx}% Include figure files
\usepackage{dcolumn}% Align table columns on decimal point
\usepackage{bm}% bold math
\usepackage{color}
\usepackage{sidecap}
\usepackage{amssymb}
\begin{document}

\title{Resilience of gated avalanche photodiodes against bright illumination attacks in quantum cryptography}
\author {Z. L. Yuan}
\email{zhiliang.yuan@crl.toshiba.co.uk}
\author {J. F. Dynes}
\author {A. J. Shields}
\address{Toshiba Research Europe Ltd, Cambridge Research Laboratory, 208 Cambridge Science Park, Milton Road, Cambridge, CB4~0GZ, UK }

\date{01 March 2011}
%\today}% It is always \today, today,

\begin{abstract}
Semiconductor avalanche photodiodes (APDs) are commonly used for single photon detection in quantum key distribution. Recently, many attacks using bright illumination have been proposed to manipulate gated InGaAs APDs. In order to devise effective counter-measures, careful analysis of these attacks must be carried out to distinguish between incorrect operation and genuine loopholes. Here, we show that correctly-operated, gated APDs are immune to continuous-wave illumination attacks, while monitoring the photocurrent for anomalously high values is a straightforward counter-measure against attacks using temporally tailored light.
\end{abstract}


\maketitle

As a solution to the key distribution problem, quantum key distribution (QKD) has attracted a great deal of interest, because its underlying security is not reliant on any assumptions about an eavesdropper's (Eve's) power. Although QKD protocols can be proven to be perfectly secure, the differences between a theoretical protocol and its real-world implementation should be carefully analyzed. This helps expose weakness of specific implementations, and thus effective counter-measures can be devised. One example of this is the decoy protocol,\cite{hwang03,wang05,lo05} which not only defeats the photon number splitting attack\cite{dusek99,brassard00} but also allows weak laser systems to achieve the highest key rates.\cite{dixon08,dixon10}

Scrutinizing the security of QKD systems using avalanche photodiodes (APDs) is important, because they are widely used.\cite{dixon10,dixon08,peev09,uqcc2010} Recently, Lyderson \textit{et al.}\cite{lydersen10} used two research QKD systems to demonstrate that continuous-wave (CW) illumination can blind InGaAs APDs so that the count rate falls to zero exactly. Under such induced blindness, Eve can gain full information about the secret key in a modified intercept-resend attack with strong resent pulses. This blinding attack would be a concern for practical QKD systems if proven universally effective. Fortunately, a later experiment showed that the blinding attack is ineffective for APDs that are operated correctly.\cite{yuan10b}

In addition to the original blinding attack, Lydersen \textit{et al.} have also proposed a broader range of illumination attacks targeting gated detectors, including thermal blinding, thermal blinding of frames, and ``sink-hole" attacks.\cite{lydersen10b} These attacks need to be carefully analyzed. In particular, efforts must be made to distinguish between incorrect operation and genuine loopholes, only after which counter-measures can be effectively constructed.  Here, we study the behavior of gated InGaAs APDs under illumination ranging from 1~fW to $>$10~mW. Careful analysis reveals that their gain modulation naturally fends off CW bright illumination attacks, without resorting to further countermeasures. Photocurrent monitoring is effective to foil more sophisticated attacks involving temporally tailored illumination.

\begin{figure}[b]
%\newpage
\centering
\includegraphics[width=.8\columnwidth]{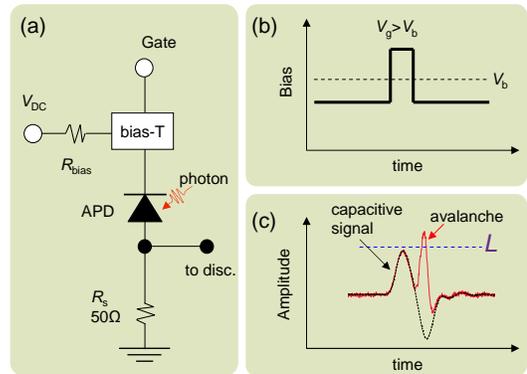}
\caption{(a) A schematic diagram for gated mode operation; (b) APD gating; (c) APD output waveforms showing capacitive responses and a single photon-induced avalanche. $L$: discrimination level.}
\label{fig:schematic}
\end{figure}

Figure~1(a) shows a typical circuit for a gated APD. The biasing resistor ($R_{bias}$) is redundant, included here only for illustrating the blinding attack.\cite{lydersen10} A voltage pulse is used to raise the APD bias ($V_g$) above its breakdown voltage ($V_b$) [Fig. 1(b)].  Under such excessive bias ($V_{ex}=V_g-V_b>0$), an APD can probabilistically multiply a single-photon induced charge into a macroscopic current. A detection event is registered when the voltage drop across the sensing resistor ($R_s$) exceeds the discrimination level ($L$), as illustrated in Fig. 1(c). It is common and good practice to set $L$ as low as possible, determined here by the capacitive response. A much lower $L$ can be achieved when the capacitive signal is removed.\cite{bethune00,tomita02,namekata06,yuan07}

The APDs used here have absolute maximum ratings of 3~mW for optical illumination and 5~mA for reverse current. However, to investigate fully bright illumination of APDs, we exceeded these ratings by manyfold, which inevitably caused damage to the APDs. As a result, two APDs were used to complete this study. Unless explicitly stated, the data shown in this paper refer to APD1. We adopt the same electrical setup as in our QKD experiment.\cite{gobby04}  The DC voltage is set 1.5~V below $V_b$. The gating pulse has 3.5~ns duration, 4~V amplitude ($V_{ex}$=2.5~V) and 2~MHz repetition rate. Electrically cooled to --30$^\circ$C, both APDs were measured to have a single photon detection efficiency of 11\%; a value typical for InGaAs detectors. A CW laser of 1550-nm is used to illuminate the APD. %The DC current is monitored.

\begin{figure}[]
\newpage
\centering
\includegraphics[width=.8\columnwidth]{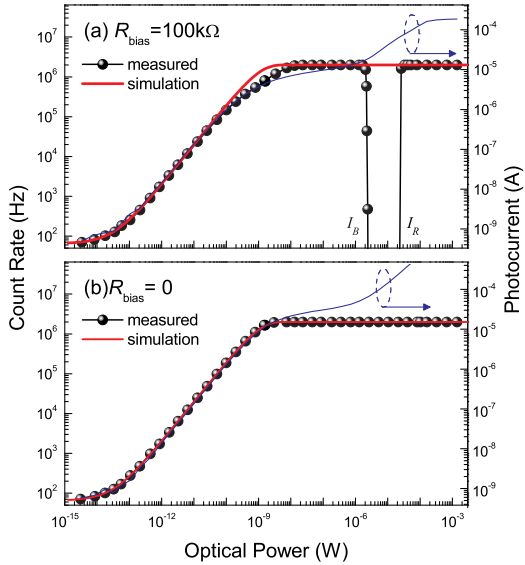}
\caption{Measured (symbols) and simulated count rates and photocurrent $vs$. incident optical power for (a) $R_{bias}=100$~k$\Omega$ and (b) $R_{bias}=0$.}
\label{fig:quenching}
\end{figure}

We first illustrate the detector blinding by use of a redundant $R_{bias}=100$~k$\Omega$.
Figure~2(a) shows the dependencies of the count rate and photocurrent as a function of the illumination power. Both are found to have a linear dependence for an optical power of 1~nW or less. Further increasing the illumination intensity, the count rate first saturates, then falls sharply to zero, and remains at zero until it finally recovers abruptly to the saturated rate. The transitions are measured at $I_B=2.5$ and $I_R=26$~$\mu$W for the count rate fall and recovery, respectively. The APD can be blinded by illumination at at any power within the zero-count gap, and then manipulated using short optical pulses.\cite{lydersen10}

The zero-count gap in Fig. 2(a) is simply a consequence of the high impedance bias resistor. Removing $R_{bias}$ causes the zero-count gap to disappear throughout the optical power range completely, as shown in Fig. 2(b). Without such a gap, Eve cannot manipulate this detector and hence the APD is secure.
The count rate is well simulated with
\begin{equation}
    R = f_0[(1-e^{-\mu \eta})+ P_d-(1-e^{-\mu\eta})\cdot P_d)],
\end{equation}
\noindent where $f_0$ is the gate repetition rate, $\mu$ the photon flux within each detection gate, $\eta$ the detection efficiency, and $P_d$ is the dark count probability. In contrast, the simulation for $R_{bias}=100$~k$\Omega$ gives a higher count rate than measured when approaching count rate saturation around 1~nW, suggesting a decreasing photon detection efficiency due to the APD bias reduction via the high impedance resistor.

The zero-count gap in Fig.~2(a) is a result of the bright illumination induced photocurrent causing a voltage drop across $R_{bias}$ and thus reducing the bias applied to the APD.  At $I_B=2.5$~$\mu$W, the photocurrent is measured as 19~$\mu$A, corresponding to a voltage drop of 1.9~V across the $R_{bias}$.  Although this voltage drop is smaller than $V_{ex}$, it is sufficient to prevent the avalanche from evolving above the discrimination level.\cite{yuan10c} The count recovery at $I_R$ is due to gain modulation, which is discussed later.

We stress that $R_{bias}=0$ is not necessary to avoid the zero-count gap. Figure~3 shows the $R_{bias}$ dependence of $I_B$ and $I_R$. With decreasing $R_{bias}$, the zero-count gap narrows, as $I_B$ increases more rapidly than $I_R$. It is determined that for $R_{bias}\leq 20$~k$\Omega$, no zero-count gap is found: the APD behaves similar to the case of $R_{bias}=0$ [Fig.~2(b)]. 

\begin{figure}[]
\newpage
%\centering
\includegraphics[width=.9\columnwidth]{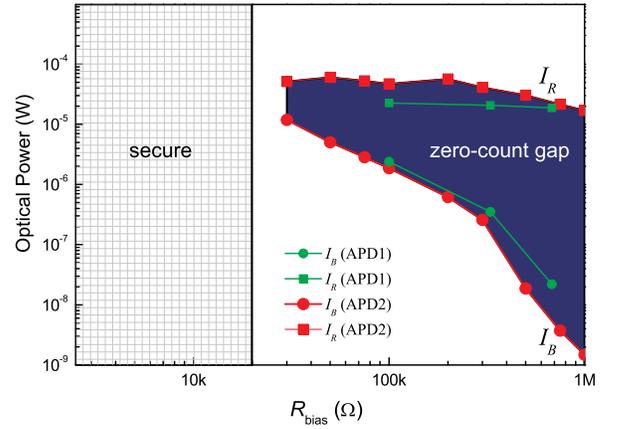}
\caption{$I_B$ and $I_R$ as a function of $R_{bias}$.}
\label{fig:quenching}
\end{figure}

Setting an inappropriate discrimination level also leads to the detrimental zero-count gap. To illustrate this, we plot in Fig.~4 the AC amplitude of the APD output as a function of illumination power for various values of $R_{bias}$. As the illumination power increases, the AC amplitude declines first, then varies slowly, and finally recovers sharply.  While all AC voltages are greater than the capacitive signal, the amplitude minimum increases with decreasing $R_{bias}$. For $R_{bias}\leq5$~k$\Omega$, the minimum is significantly greater than the capacitive signal with clear discrimination from the capacitive signal, thus allowing persistent counting. However, if we deliberately discriminate at a level which is twice the capacitive signal, a detrimental zero-count gap would emerge even for the case of $R_{bias}=0$.

The AC amplitudes shown in Fig.~4 are provided by the photocurrent modulated by the APD gating. In contrast to single photon detection, this gain modulation signal becomes significant only with bright illumination. The sharp amplitude recovery at high powers is due to the APD bias reaching to its punch-through voltage, below which photon absorption does not produce a photocurrent.\cite{jiang07} At this voltage, electrical gating switches on/off the photocurrent, thus producing a large pulsed current.

\begin{figure}[t]
\newpage
\centering
\includegraphics[width=.9\columnwidth]{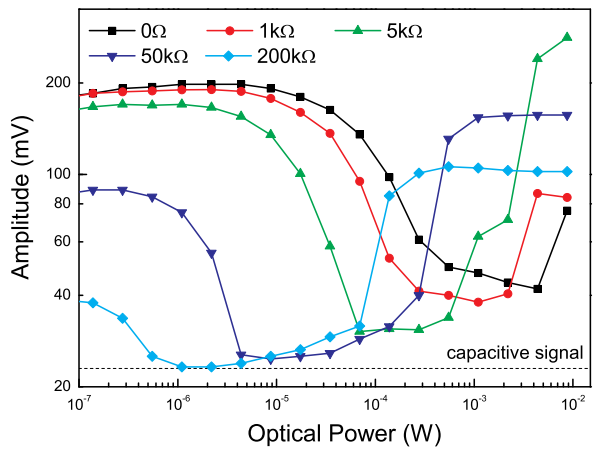}
\caption{Signal amplitudes by gain modulation as a function of illumination power for varying $R_{bias}$. The data were obtained from APD2.}
\label{fig:r1k}
\end{figure}

Gain modulation might be compensated by a sophisticated Eve using temporally tailored illumination.\cite{lydersen10b} In this case, monitoring the photocurrent for anomalously high values is a straightforward counter-measure.  As shown in Fig.~2, the photocurrent is proportional to the count rate in the single photon regime ($I<1$~nW). Such a relation is also observed for high speed APDs.\cite{yuan09,yuan10} By constantly monitoring the photocurrent, an anomalously high measured photocurrent, \textit{i.e.}, exceeding the expected photocurrent in the single photon regime, can be used to foil any bright illumination attack. We stress that this measure is effective universally for all APDs, including the non-gated\cite{makarov09} as well as high-speed gated.\cite{yuan07,namekata06,thomas10} In comparison with the previous proposal using a separate power-meter,\cite{makarov09} the present solution does not need a lossy beam splitter which will inevitably deteriorate the bit rate and distance in QKD.

Before commenting on recent attacks, two measures for safeguarding APDs are listed below:
\begin{enumerate}
  \item Avoid a high impedance biasing resistor and use a low discrimination level;
  \item Monitor the photocurrent for anomalously high values.
\end{enumerate}
\noindent  Measure 1 is actually the very basic rule for correctly operating a gated APD.\cite{yuan10b} Following this rule alone will prevent CW bright illumination attacks.

Now, we discuss what has caused the vulnerability in Clavis2, one of the two systems attacked by Lydersen \textit{et al.}\cite{lydersen10}  We choose this system for discussion simply because it has more experimental details\cite{lydersen10b} available. While APDs in Clavis2 have an adequate $R_{bias}=1$~k$\Omega$, their discrimination levels were set at $\sim$80~mV, which is more than twice the value needed for rejection of the capacitive signal (35~mV).\cite{lydersen10b} Using such high discrimination level, induced blindness was then demonstrated at an illumination of 397~$\mu$W. Later, instead of correcting the discrimination levels, Lydersen \textit{et al.} removed the non-zero $R_{bias}$ to show the induced blindness once again at a much higher power ($\sim$10~mW).\cite{lydersen10b}
This time, the detector heating was suggested as the prime cause.
In both cases, we believe the induced blindness should have been avoided if the discrimination levels were correctly set.
No matter whether the illumination causes APD bias reduction\cite{lydersen10} or device heating,\cite{lydersen10b} gain modulation always exists and should trigger a discrimination level that is correctly set.
In our tests on an APD detector from Clavis2's manufacturer, we saw no evidence of an induced blindness for a CW illumination of up to 14~mW.\cite{id200} Nor did we see the induced blindness for an illumination of up to 17.8~mW on our own detector.\cite{yuan10b}

Lydersen \textit{et al.} also demonstrated an alternative thermal attack using temporally tailored light.\cite{lydersen10b}
Attacking the Clavis2, which transmits quantum signals in frames, blinding illuminations are switched on only during the intervals between QKD frames.  First, this is not a stealth attack. It gives itself away by causing extra photon clicks outside QKD frames.
Even if a QKD system ignores such clicks, monitoring the photocurrent for anomalously high value is sufficient to foil this attack.  The attack still requires an average optical power of 1.5~mW, the photocurrent of which will significantly exceeds the value expected in the single photon counting regime.

As a more special attack targeting AC-coupled detectors, ``sink-hole" attack illuminates between gates, creating a photocurrent valley around each APD gate.\cite{lydersen10b} As AC coupling re-bases the ground level, the avalanche signal sitting in these valleys will be reduced below the discrimination level, thereby preventing detection of single photons.  However, this attack still requires an illumination of around 200~$\mu$W, the photocurrent of which remains easily detectable. Moreover, such sink-hole attack is ineffective to detectors using DC coupling, in which no capacitor is used to block the DC signal before the discriminator.

Finally, we discuss the ``after-gate" attack, in which Eve exploits the intrinsic linear mode of APD between gates, facilitated by the excessively long modulation gates or detection acceptance window in the QKD setup.\cite{wiechers11} As detailed in Ref.~[{\onlinecite{wiechers11}], this attack induces high levels of detector afterpulsing, and thus works only on frame-based QKD systems with long intervals between frames. This attack is ineffective on continuous-running one-way systems, particularly the high speed ones.\cite{dixon08,dixon10} Moreover, for any QKD systems, either narrowing the modulation gate duration or the detection acceptance window will completely defuse this attack. Nowadays, use of sub-nanosecond modulation pulses or acceptance windows in QKD is not uncommon.\cite{dixon08,dixon10,uqcc2010,tang08}

To conclude, correctly-operated, gated APDs are immune to CW bright illumination attacks.  For temporally tailored illumination, monitoring the photocurrent for anomalously high values is a straightforward countermeasure.

Partial support from EU project QEssence is acknowledged.

%\newpage

%\bibliography{GatedAPD}% Produces the bibliography via BibTeX.

\begin{thebibliography}{26}
\bibitem [{id2()}]{id200}%
  \BibitemOpen
  \href@noop {} {}\bibinfo {note} {Model: id200. Settings: 4~MHz gating
  frequency and 2.5~ns gate duration. The discrimination level, as well as the
  bias impedance, is unknown to end-users.}\BibitemShut {Stop}%
\end{thebibliography}


%\newpage

\end{document}